\documentclass[11pt]{article}
\usepackage{graphicx,graphics}
\usepackage{epstopdf}
\usepackage[a4paper,hmarginratio=1:1,vmarginratio=2:3,totalwidth=15.2cm,totalheight=22.7cm]{geometry}
\usepackage{bm,epstopdf,epsfig,amsmath,amssymb,amsfonts,colordvi,latexsym,comment,cancel,
verbatim,
slashed}
\usepackage[font=md,captionskip=8pt]{subfig}
\usepackage[usenames,dvipsnames]{color}

\usepackage{setspace}

\usepackage[utf8]{inputenc}

\newcommand{\cA}{{\cal A}}

\newcommand{\cO}{{\cal O}}

\newcommand{\Tr}{\mbox{Tr}}

\newcommand{\ra}{\rightarrow}
\newcommand{\be}{\begin{equation}}
\newcommand{\ee}{\end{equation}}
\newcommand{\bea}{\begin{eqnarray}}
\newcommand{\eea}{\end{eqnarray}}
\newcommand{\Ra}{\Rightarrow}

\newcommand{\otheta}{\overline\theta}

\newcommand{\baa}{\begin{array}}
\newcommand{\eaa}{\end{array}}
\long\def\symbolfootnote[#1]#2{\begingroup
\def\thefootnote{\fnsymbol{footnote}}\footnote[#1]{#2}\endgroup}
\begin{document} 
\begin{flushright}
CERN-TH-2017-072\\
\end{flushright}

\bigskip\medskip

\thispagestyle{empty}

\vspace{3.5cm}

\begin{center}
  {\Large {\bf 
  Wilson lines and UV sensitivity
\bigskip
in magnetic compactifications}}

\vspace{1.cm}

 {\bf D. M. Ghilencea}$^{\,a}$, \,\,  {\bf Hyun Min Lee}$^{\,b}$
\symbolfootnote[1]{E-mail: dumitru.ghilencea@cern.ch, hminlee@cau.ac.kr}
\bigskip

$^a$ {\small Theoretical Physics Department,
 National Institute of Physics 

and Nuclear Engineering  Bucharest, MG-6 077125  Romania}

$^b$ {\small Department of Physics, Chung-Ang University, 06974 Seoul, Korea}

\end{center}

\begin{abstract}
We investigate the ultraviolet (UV) behaviour of 6D N=1 supersymmetric effective (Abelian)
 gauge theories  compactified on a two-torus ($T_2$) with magnetic flux. 
To this purpose  we compute offshell
 the one-loop correction to the   Wilson line state self-energy. The  offshell calculation 
is actually {\it  necessary}  to capture the usual effective field theory expansion 
in powers of $(\partial/\Lambda)$.  Particular care is paid to  the regularization of the 
(divergent)  momentum integrals, which is relevant for  identifying the corresponding counterterm(s).
We find a  counterterm which is a new higher dimensional effective operator of dimension d=6, that is
enhanced for  a larger  compactification area (where the effective theory applies)
and is consistent with the symmetries of the theory. 
Its consequences are briefly discussed and comparison is made with orbifold 
compactifications without flux.
\end{abstract}

\newpage

\section{Introduction}

In this letter we explore the ultraviolet behaviour of  supersymmetric 
models compactified to four dimensions on a two-torus $T_2$ in the presence of magnetic flux.
Compactifications  on tori with magnetic fluxes were investigated   in
 string theory  e.g. \cite{Berkooz}-\cite{Cremades} (for a review \cite{Ralph,Carlos}) and
are interesting because they can break supersymmetry and lead to chiral fermions 
\cite{Bachas}.  This motivated the interest in effective theory approach to model 
 building e.g. \cite{B1}-\cite{F}.

In this work we  compute (offshell) the one-loop correction to a two-point
Green function of the self-energy of a complex scalar field $\varphi$ in a compactification  
to four dimensions of a 6D N=1 supersymmetric Abelian gauge theory on $T_2$ with magnetic flux.
The scalar field $\varphi$
is actually a Wilson line state, which is a  fluctuation 
of a combination of components $A_{5,6}$  of the gauge fields $A_M$ ($M\!=\!\mu,5,6$). 
The motivation is two-fold:
 few quantum investigations exist for such compactification
and   the field $\varphi$  may play the role of a higgs field in realistic models,
which is interesting for model building and the hierarchy problem.

We pay particular attention to the regularization
of the quantum corrections. Indeed, the one-loop integrals are divergent and call
for a UV regularization consistent with the symmetries of the theory. 
The regularization  ensures that (the series of) these integrals are well-defined and
any divergences  of the result in the limit of removing the regulator
dictate  the form of the corresponding counterterm operators.  Since 
 effective theories are non-renormalizable, the counterterms may be higher dimensional
operators. 
The {\it offshell} calculation is important and is actually {\it necessary} in order to 
capture the behaviour of the effective theory which is an expansion in powers\footnote{
Most quantum corrections from  extra dimensions are computed onshell. For offshell 
results see \cite{G1,G2}.}
 of $(\partial/\Lambda)$ \cite{georgi}
 where  $\Lambda$ is a high  scale (e.g. compactification scale).
We use dimensional regularization (DR), 
since it respects all symmetries, in particular gauge symmetry.
We  then  compare the UV behaviour of our result for the quantum correction in the presence of
magnetic flux against similar  results in orbifold compactifications without flux
such as effective theory on   $T_2/Z_2$, etc.

\section{Magnetic compactification on a torus}

We begin our study with the relevant part of the
 action. Consider first the action of a
  6D N=1 vector superfield and hypermultiplet compactified to 
4D on a torus $T_2$ in the presence of magnetic flux. This can 
be described in 4D N=1 superfields language \cite{M}.
For the  details of this compactification we refer the reader to
\cite{B1,I,H,t1}. 
For the vector superfield
\bea
S_v&=&
\!\int d^6x \Big\{
\int d^4\theta \Big[\partial V\overline\partial V+\Phi^\dagger\Phi+\sqrt 2 \,V\,
(\overline\partial \Phi+\textrm{h.c.}) \Big]
+
\frac14 \int d^2\theta \, W^\alpha W_\alpha +\textrm{h.c.}
\Big\}
\eea

\medskip\noindent
where $\partial\equiv \partial_5-i \partial_6$. Only zero-modes of $V$ (hereafter $V_0$), of
gauge kinetic field-strength $W$ ($W_0$)  and of
the superfield $\Phi$ ($\Phi_0$) are relevant below. We have
$\Phi_0\vert_{\theta=\otheta=0}=1/\sqrt 2 (A_6+iA_5)+\varphi$, where $\varphi$ 
defines  a complex continuous Wilson line state on $T_2$.

We also need  a 6D N=1 hypermultiplet of chiral superfields $Q$, $\tilde Q$ 
of  charges $\pm q_0$
\medskip
\bea
S_h=\!\!\int\! d^6x \Big\{\!
\int\! d^4\theta\,\Big[ Q^\dagger e^{2\,q_0\,g\,V}\,Q+\tilde Q^\dagger e^{-2\,q_0\,g\,V}\,\tilde Q\Big]
\!+
\Big[
\int\! d^2\theta\,\,\tilde Q \,(\partial +\sqrt 2 g\,q_0\,\Phi )\,Q+\textrm{h.c.}\Big]
\Big\}
\eea

\medskip\noindent
with $g$ the gauge coupling.
One must integrate $S_h,S_v$ over $T_2$ in the presence of magnetic flux \cite{B1},
but a set of  basis functions is required. First, we
 use a symmetric gauge choice with
$A_5=(- 1/2) f x_6$ and $A_6=(1/2) f x_5$ 
($f$=constant), satisfying  a constant field strength 
$F_{56}=\partial_5 A_6-\partial_6 A_5=f$.
Its flux through $T_2$ closed surface
 is then quantised\footnote{Another way to see the quantisation
 condition is the following. We can make a gauge choice near
 $x_5=0$ and $x_5=2\pi R_5$: Region I ($-\pi R_5 <x_5< \pi R_5$): 
$A_5=0,  A_6 = f x_5$, and Region II ($\pi R_5< x_5< 2\pi R_5$):  $A_5=0, A_6=f (x_5 - 2\pi R_5)$.
Then, two gauge potentials are connected by a gauge transformation in 
the overlapping region:  $A_{II} - A_I = -2\pi  f  R_5 = \partial \Lambda$,
 with $\Lambda = - 2 \pi f R_5 x_6$. As a result, the wavefunctions of charged 
fields, $\phi$, are connected in this overlapping region 
as $\phi_{II} = e^{-i q_0 g f 2\pi R_5 x_6} \phi_I$. Then, single-valuedness of wavefunctions
 along the $x_6$ direction requires $q_0 g f (2\pi) R_5 R_6 = N$  with $N$ integer.
 The periodicity along the $x_6$ direction is guaranteed by the same quantisation condition. }  
$q_0 g/(2\pi)\int_{T^2} F_{56} dx_5 dx_6\!=\!q_0 g f{\cA}/(2\pi)\!\in\!Z$,
($\cA$ is the area of the torus).
The Kaluza-Klein (KK)
spectrum of the charged fields will then resemble that of Landau levels \cite{B1,I,H,Landau}.

To find the basis set of functions, notice that
covariant derivatives $D_k=\partial_k+i q_0 g A_k$ ($k=5,6$) satisfy 
 $[iD_5,i D_6]= -i q_0 g f$. Assuming $f\!<\!0$,  one can construct 
a 1D harmonic oscillator Hamiltonian $H\!=\!p^2/(2m)\!+1/2\, m \omega^2 x^2$ 
of  $p\!\sim\! iD_6$ and $x\sim iD_5$, $m=1/2$, $\omega=2$. Its
 eigenfunctions define the basis set of functions $\psi_{n,j}$ \cite{B1,I,H}.
The ladder operators are  $a=(1/\sqrt\alpha)  \,(iD_5-D_6)$,
$a^\dagger=1/\sqrt \alpha \,(iD_5+D_6)$ with $[a,a^\dagger]=1$, so
 $H=\alpha (a^\dagger a+1/2)$ and 
\bea
\alpha=-2 q_0 g f= \frac{4\pi N}{\cA}>0,\qquad (N\in Z_+)
\eea
The basis functions are 
$\psi_{n,j}=(a^\dagger)^n/\sqrt{n!}\, \psi_{0,j}$, where $n$ refers to the Landau level and
$j$ reflects the $N$-fold degeneracy. These are orthonormal on $T_2$, and
 $a^\dagger\psi_{n,j}=\sqrt{n+1}\,\psi_{n,j}$, with $\psi_{0,j}$ as zero mode: $a\,\psi_{0,j}=0$.
Then  $\partial+\sqrt 2 q_0 g \, \Phi_0 =
-i \sqrt{\alpha} \, a^\dagger +\sqrt 2 q_0 g\, \varphi$, which
 is used in  $S_h$,  
together with an expansion of superfields in the basis functions $\psi_{n,j}(x_m)$:
\medskip
\bea
Q(x_M,\theta,\otheta)=\sum_{n,j} Q_{n,j}(x_\mu,\theta,\otheta)\,\psi_{n,j} (x_m),\qquad M=\mu, 5, 6.
\eea
A similar expansion exists for $\tilde Q(x_M,\theta,\otheta)$ in this basis, with 
coefficients $\tilde Q_{\tilde n,\tilde j}(x_\mu,\theta,\otheta)$.
One finds the relevant part of the 4D action \cite{B1}
\medskip\bea\label{act}
S&\supset&\int d^4x\Big\{
\int d^4\theta \,\Big[\varphi^\dagger\varphi 
+\sum_{n,j} Q_{n,j}^\dagger e^{2\,q_0\,g\,V_0} Q_{n,j}
+\sum_{n,j} \tilde Q_{n,j}^\dagger e^{- 2\,q_0\,g\,V_0} \tilde Q_{n,j} 
+ 2 f V_0\Big]
\nonumber\\[-3pt]
&+&\!\!\!\!\int\! d^2\theta\,
\Big[\frac14 
W_0^\alpha W_{0,\alpha} -i \sum_{n,j} \sqrt{\alpha (n+1)}\, \tilde Q_{n+1,j} Q_{n,j} 
+\sum_{n,j}\sqrt{2}\, q_0\, g\,\tilde Q_{n,j}\,\varphi\,Q_{n,j} \Big]\!+\textrm{h.c.}
\Big\}\,\,\,\,\,\,
\eea

\medskip\noindent
where we kept only the zero modes of the gauge kinetic term  and of Wilson line scalar $\varphi$.
After eliminating the auxiliary fields one identifies
the scalar fields mass:  $m^2_{\tilde Q_{n,j}}=m^2_{Q_{n,j}}=\alpha (n+1/2)$;
for  fermions their mass  can be read  from the last line of the above equation:
 $m^2_{\Psi_{n,j}}=\alpha (n+1)$ for a Dirac fermion composed of two Weyl spinors as in 
 $\Psi_{n,j}\equiv ({\tilde\chi}_{n+1,j},\chi_{n,j})^T$. 
The (onshell-SUSY) couplings of these fields, in  components, are:
\medskip
\bea
L&=&-i \sqrt 2 q_0 g \sum_{n,j} \sqrt{\alpha (n+1)}\, \varphi 
\big[\tilde Q^\dagger_{n+1,j} \tilde Q_{n,j}- Q_{n,j}^\dagger Q_{n+1,j}\big]
-\sqrt 2 q_0 \,g\,\sum_{n,j} \varphi\,\tilde\chi_{n,j}\chi_{n,j}
+\textrm{h.c.}
\nonumber\\
&-& 2q_0^2\,g^2 \sum_{n,j} \big[ \vert Q_{n,j}\vert^2+\vert \tilde Q_{n,j}\vert^2\big]\,\vert\varphi\vert^2
\eea
%
where the sums are over $n\geq 0$;
 $\tilde\chi$ ($\chi$) are the Weyl spinors of $\tilde Q$ ($Q$) superfields.
With this information we can  investigate the quantum corrections to the
mass of the scalar field $\varphi$.

\section{One-loop corrections to  Wilson line}

\begin{figure}[t!] 
\begin{center}
\begin{tabular}{cc|cr|} 
{\psfig{figure=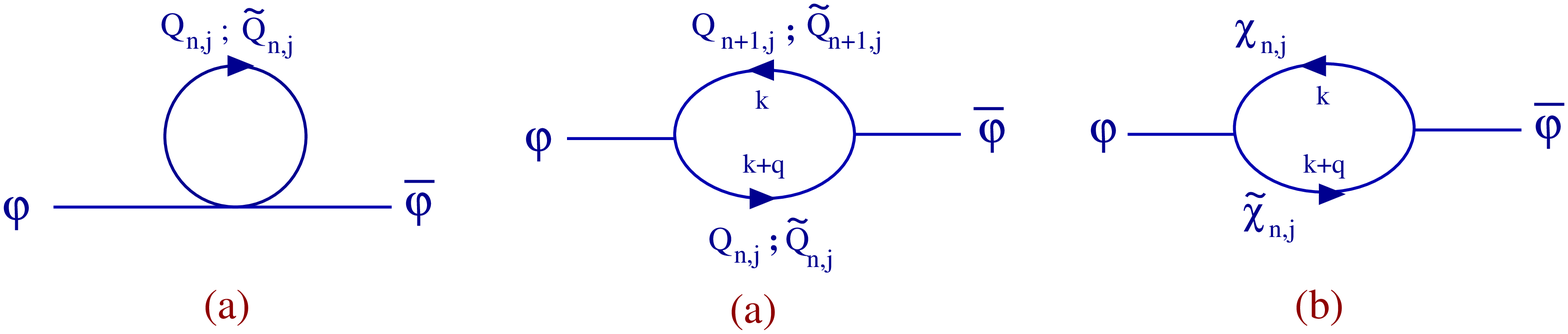,
 height=3.cm,width=14cm}} 
\end{tabular}%
\end{center}
\def\baselinestretch{1.}
\vspace{-0.3cm}
\caption{\small One-loop diagrams, at external momentum $q$,
of  bosonic (a) and fermionic (b) contributions, respectively, to 
 the Wilson line scalar ($\varphi$) self-energy.}
\label{oneloop}
\end{figure}

With the above action, 
we compute the one-loop corrections to the Wilson line scalar, shown in fig.\ref{oneloop}
for non-vanishing external 4-momentum $q$. This allows us to investigate their UV behaviour
under scaling of the momentum. Since the integrals are divergent,
we use the DR scheme, in order to find the poles
and identify their corresponding counterterms. This regularization preserves all symmetries 
of the theory. After performing a Wick rotation to
the Euclidean space and with the  DR subtraction scale $\mu$ introduced to ensure
dimensionless coupling ($g$) in $d=4-2\epsilon$ dimensions, we find for the bosonic 
contribution 
\medskip
\bea
\delta m_b^2(q^2)=2\, q_0^2\, g^2\, N \mu^{2\epsilon}
\sum_{n\geq 0} \int \frac{d^d k}{(2\pi)^d}\frac{2 \,k^2+\alpha}{
\big[ (q+k)^2+\alpha\,(n+1/2) \big]\big[k^2+\alpha \,(n+3/2)\big]}.
\label{b}
\eea
For the fermionic part
\bea
\delta m_f^2(q^2)=-2\, q_0^2\, g^2\, N \mu^{2\epsilon}\sum_{n\geq 0} \int \frac{d^dk}{(2\pi)^d}
\frac{2\,k\,(q+k)}{\big[ (q+k)^2+\alpha\,n\big]\,\big[ (k^2+\alpha\,(n+1)\big]}.\qquad\quad
\label{f}
\eea

\medskip
\noindent
Performing the integrals in the DR scheme (see  the Appendix) gives\footnote{
Unlike in 6D orbifolds, in the present case only one KK sum is present,
 which would apparently  make the result less UV  divergent. This is however
 misleading because  in the present case the (masses)$^2$ under the sum are  
linear rather than quadratic in the level ($n$), 
thus there is no  UV improvement in this sense.}
\medskip
\bea
\delta m_b^2(q^2)\!\!\!&=&\!\!\! K_0\, (4\pi\mu^2/\alpha)^\epsilon \int_0^1 \!\! dx 
\Big[ (2 q^2\,x^2+\alpha)\, \Gamma[\epsilon]\,\zeta[\epsilon,\rho_1] 
+ 
d \,\alpha\,\Gamma[-1+\epsilon] \,\zeta[-1+\epsilon,\rho_1]\Big]
\nonumber\\
\delta m_f^2(q^2)\!\!\!&=&\!\!\!\! - K_0\, (4\pi\mu^2/\alpha)^\epsilon\! \int_0^1 \!\! dx 
\Big[ 2 q^2\,x (x-1)\, \Gamma[\epsilon]\,\zeta[\epsilon,\rho_2] 
+ 
d\,\alpha\,\Gamma[-1+\epsilon] \,\zeta[-1+\epsilon,\rho_2]\Big]
\,\,\,\label{q1}
\eea

\medskip
\noindent
with the notation
\bea
K_0\equiv \frac{2 q_0^2 g^2 N}{(4\pi)^2},\qquad
\rho_2=\rho_1-\frac12
=(1-x) \Big(1+ x \,\frac{q^2}{\alpha}\Big)>0
\eea

\medskip
\noindent
where we introduced the Hurwitz zeta function  $\zeta[s,a]=\sum_{n\geq 0} (n+a)^{-s}$
 \cite{E,R}.
The above bosonic and fermionic contributions have poles from Gamma functions, $\Gamma[\epsilon]$ and
$\Gamma[-1+\epsilon]$. 
One could proceed in eqs.(\ref{q1})  to Taylor expand the zeta functions for 
small $\epsilon$ and isolate the poles from the finite part, however,  one 
cannot then integrate the resulting  terms involving
 $(d/d z\, \zeta[z,\rho])_{z=-1}$ since for this derivative 
only asymptotic expansions are known \cite{E}. To avoid this, we integrate by parts
 the second term in both  $\delta m_{b,f}^2$ and use
\medskip
\bea
\frac{\partial \zeta[s,\rho]}{\partial \rho}=-s \,\zeta[s+1,\rho]
\eea
This gives
\bea
\delta m_b^2(q^2)\!\!\!&=&\!\!\!K_0 
\Big(\frac{4\pi\mu^2}{\alpha}\Big)^\epsilon \Big[
d\,\alpha\, \Gamma[\epsilon-1] \,\zeta[-1+\epsilon,1/2]
+
\Gamma[\epsilon]\!\int_0^1\!\! dx\,  \zeta[\epsilon,\rho_1]\, f_1(x)
\,\Big]
\nonumber\\
\delta m_f^2(q^2)\!\!\!&=&\!\!\!\! -K_0 
\Big(\frac{4\pi\mu^2}{\alpha}\Big)^\epsilon \Big[
d\,\alpha\, \Gamma[\epsilon-1] \,\zeta[-1+\epsilon,0]
+
\Gamma[\epsilon]\!\int_0^1\!\! dx\,  \zeta[\epsilon,\rho_2]\, f_2(x)
\,\Big]\nonumber
\eea

\medskip
\noindent
Here
$f_1(x)=2 \,q^2\,x^2 +\alpha +x \,d\,\alpha \rho_1^\prime(x)$ 
and
$f_2(x)=2 \,q^2\,x (x-1) +x \,d\,\alpha \rho_2^\prime(x)$
with a notation
$\rho_j^\prime(x)=(d/d x) \rho_j(x)$, $j=1,2$. Further
\bea
\Gamma[\epsilon]&=&\frac{1}{\epsilon}-\gamma_E+\cO(\epsilon)
\nonumber\\
\zeta[\epsilon,\rho_j]&=&\zeta[0,\rho_j]+\epsilon\,\zeta^{(1,0)}[0,\rho_j]+\cO(\epsilon^2),\quad j=1,2.
\nonumber\\
\Big(\frac{4\pi\mu^2}{\alpha}\Big)^\epsilon&=&1+\epsilon\ln\frac{4\pi\mu^2}{\alpha}+\cO(\epsilon^2)
\eea
with the Euler constant 
$\gamma_E\approx 0.577216$. We 
then find\footnote{One has $\zeta[0,\rho]=1/2-\rho$,
$\zeta^{(1,0)}[0,\rho]=\ln\Gamma[\rho] -\ln\sqrt{2\pi}$, 
and $\zeta^{(1,0)}[-1,\frac{1}{2}]=-\frac{1}{2}\zeta^{(1,0)}[-1,0]-
\frac{1}{24}\ln 2$  \cite{R}.}
\medskip
\bea
\delta m_b^2(q^2)&=&K_0\Big[
-\frac{q^4}{30\,\alpha} \,\Big(\frac{1}{\epsilon}+\ln\frac{4\pi\mu^2e^{-\gamma_E}}{\alpha}\Big)
-\frac{\alpha}{12} \ln\frac{e^3\,G^{24}}{4}
-\frac{q^2}{6}
-\frac{q^4}{30\,\alpha} 
+ H_1(q)\Big]+\cO(\epsilon)
\nonumber\\
\delta  m_f^2(q^2)&=&\!\!\!\! -K_0 \Big[\,\,\,\frac{4 \,q^4}{30 \,\alpha}\,\,\Big(\frac{1}{\epsilon}
+\ln\frac{4\pi\mu^2 e^{-\gamma_E}}{\alpha}\Big) 
+\alpha\ln G^4
-\frac{q^4}{30\,\alpha}
+H_2(q)\Big]+\cO(\epsilon)
\label{gg}
\eea

\medskip\noindent
with  $G=1.28243$  the Glaisher constant\footnote{
Glaisher constant is given by $\ln G=1/12-\zeta^\prime[-1]$, with $\zeta[x]$ the Riemann 
zeta function.}$^,$\footnote{
The poles in  eqs.(\ref{gg})  are identical to those obtained if we Taylor expanded
the expressions in eqs.(\ref{q1}) 
about $\epsilon=0$, and used $\zeta[0,\rho]=1/2-\rho$, and $\zeta[-1,\rho]=
-1/2\, (\rho^2-\rho+1/6)$ \cite{R}.}.
Above we introduced the functions $H_1$, $H_2$
\medskip
\bea
H_1(q)&=&
\int_0^1 dx\, (2\, q^2\, x^2 
+\alpha +4\, x\, \alpha\, \rho_1^\prime(x)) \ln \frac{\Gamma[\rho_1(x)]}{\sqrt {2\pi}}
\nonumber\\
&=&
\frac{\alpha}{12}\ln\frac{G^{24}\,e^3}{4}
+\frac{q^2}{6}
-\frac{9\,\zeta[3]}{8\pi^2}\,q^2
+c_b\,\frac{q^4}{\alpha}+\cO((q^2/\alpha)^3)
\eea

\medskip\noindent
with 
 $c_b=-109/720-(1/15) \ln 2+ \ln G -14\,\zeta^\prime[-3])\approx -0.02414$, and
\medskip
\bea
H_2(q)&=&
\int_0^1 dx\, (2 q^2\,x\,(x-1)+4\alpha \,x\,\rho_2^\prime(x))\ln \frac{\Gamma[\rho_2(x)]}{\sqrt{2\pi}} 
\nonumber\\
&=& 
-\alpha\ln G^4+\frac{3\zeta[3]}{2\pi^2}\,q^2+ c_f\,\frac{q^4}{\alpha}+\cO((q^2/\alpha)^3)
\eea

\medskip\noindent
and $c_f=11/45+16\, \zeta^\prime[-3]\approx 0.3305$.
Therefore, up to irrelevant $\cO(\epsilon)$ terms
\bea
\delta m_b^2(q^2)& = & K_0 \Big[-\frac{q^4}{30\alpha\epsilon}+ \cO(q^2/\alpha)\Big]
\nonumber\\
\delta m_b^2(q^2)& = & K_0 \Big[-\frac{4 q^4}{30 \alpha\epsilon}+ \cO(q^2/\alpha)\Big]
\eea

\medskip
\noindent
The sum of bosonic and fermionic contributions 
$\delta m^2(q^2)=\delta m_b^2(q^2)+\delta m^2_f(q^2)$, is found in general
 from eq.(\ref{gg}), but for small momenta $q^2\ll \alpha$ it simplifies 
\medskip
\bea\label{sum}
\delta m^2(q^2)=
K_0\Big[ -\frac{q^4}{6 \alpha}\,
\Big(\frac{1}{\epsilon}+\ln\frac{4\pi\mu^2 e^{-\gamma_0}}{\alpha}\Big)
 -\frac{21 \zeta[3]}{8\pi^2} q^2 +\cO(q^6/\alpha^3)\Big].
\eea

\medskip\noindent
with $\gamma_0=\gamma_E+6 (c_b-c_f)$.
A pole is present in the two-point Green function\footnote{This is a genuine 6D divergence.}, 
reflecting  the UV divergences of the theory and {\it  the limits $q^2\ra 0$ and
$\epsilon\ra 0$  do not commute}, which shows the importance
of this calculation. A finite quantum correction ($\propto q^2$) is also present.

\section{Counterterms and symmetries}

Eq.(\ref{sum})  shows that a  counterterm is needed to cancel the pole $q^4/\epsilon$.
The counterterm involves the 2-point self-energy, so it has the  form
$L_{c.t.}=-K_0/(6\alpha)\varphi^\dagger \Box^2\varphi$. 
In superfield language, this operator has the form ($\lambda$ is a new
dimensionless coupling):
\bea
L= \frac{\lambda}{\alpha}
\int d^4\theta \,\,\varphi^\dagger \Box\varphi=
-\frac{\lambda}{\alpha}
\varphi^\dagger\Box^2\varphi+\cdots
\label{ct}
\eea

\medskip\noindent
where we used the same notation for the superfield and its scalar component.

This operator respects all symmetries of the theory and its 
 presence  is  a reminder
 that our theory, although supersymmetric,  is nevertheless
 non-renormalizable. Indeed, such theories are  an expansion in powers 
 $(\partial/\Lambda)^n$ \cite{georgi}, so such counterterms are expected;
here $\Lambda$ is the scale of new 
physics (from a 4D perspective), in this case $\Lambda^2\sim \alpha\sim 1/\cA$.
Higher loops will generate more operators of this type.
This operator, often overlooked in similar quantum calculations 
due to technical difficulties,
is not specific to compactification with fluxes - 
 it  was also seen in 5D and 6D orbifold models at one-loop \cite{G1,G2}. 
The counterterm modifies  the dispersion relations (the 
poles of the propagator) of the scalar $\varphi$,  which  acquires
a new solution, ghost-like,  due  to the higher order derivative \cite{An}. 
Eqs.(\ref{sum}), (\ref{ct}) show the propagator of $\varphi$ has new pole at
\medskip
\bea
m^2_{pole}=\frac{\alpha}{\lambda}\, 
\Big[\,1+\frac{21\zeta[3]}{8\pi^2}\,K_0\Big]
\eea

\medskip\noindent
This mass state is of the order of the compactification scale $\sqrt\alpha\sim 1/\sqrt\cA$
and corresponds to the ghost degree of freedom. 
Note that the effective theory approach is reliable for 
 a large torus area/radii (or  small  flux\footnote{
One cannot take $\alpha\ra 0$ since the flux is quantised.}
 $\alpha\sim 1/\cA$) but then also operator (\ref{ct}) is enhanced!

In applications it is useful to replace this operator by an
equivalent  polynomial  one \cite{hdo1};
 this is done by a non-linear field-redefinition or, equivalently,
by using the equation of motion (in superfields) for $\varphi$:
 $-1/4 \overline D^2 \varphi^\dagger
+\sqrt 2 q_0 g \sum_{n,j}\tilde Q_{n,j} Q_{n,j}=O(\cA)$, where we used eq.(\ref{act}).
This is used back in the  action and
effectively integrates the ghost ($\overline D^2\varphi^\dagger$)
but leaves $\varphi$ in the action; then operator (\ref{ct}) 
becomes (with $-16 \varphi^\dagger\Box\varphi=\varphi^\dagger\overline D^2 D^2\varphi$)
\bea
L\propto\frac{\lambda}{\alpha} \int d^4\theta \,q_0^2 g^2
 \Big\vert \sum_{n,j} \tilde Q_{n,j} Q_{n,j}\Big\vert^2
\eea
This is a dimension-six  polynomial effective operator, equivalent
to $L$ of (\ref{ct}) and brings many non-renormalizable operators in the action,
suppressed by $\sqrt \alpha\sim 1/\sqrt \cA$.

Having identified the counterterm operator, we can now 
formally set $q^2=0$ in the one-loop correction
$\delta m_{b,f}^2(q^2)$ of eqs.(\ref{gg}) and by  using
the exact relations
\medskip
\bea
H_1(0)=(\alpha/12) \ln(G^{24} e^3/4), \qquad H_2(0)=-\alpha \ln G^4
\eea
 we find from eqs.(\ref{gg})
\bea\label{gres}
\delta m_b^2(0)=0,\qquad
\delta m_f^2(0)=0, \quad \Ra\quad
\delta m^2(0)\!\equiv \!\delta m_b^2(0)\!+\!\delta m_f^2(0)\!=\!0.
\eea

\medskip\noindent
Therefore the bosonic and fermionic contributions
vanish separately for  $q^2\ra 0$, as conjectured in  \cite{B1}.
This indicates that at one-loop  $\varphi$ is a flat direction of the corresponding
potential which has a vanishing curvature:
$\delta m^2(q^2\!\!=\!0)\!=\!0$, 
as we showed. Beyond one-loop, any quantum calculation must include  
the one-loop counterterm of eq.(\ref{ct}).

Let us comment briefly on the result of eq.(\ref{gres}).
In compactifications without flux the Wilson line $\varphi$ 
changes  the boundary conditions of the charged fields, giving  a continuous 
shift of the KK levels   masses \cite{Bachas}. Then   $\varphi$ acquires at one-loop a 
potential and  a nonzero correction  $\delta m^2(q^2\!=\!0)$  \cite{Ghilencea:2005vm}. 
By contrast, in our compactification with flux,  $\varphi=\varphi_1+i \varphi_2$
only shifts \cite{H} the argument $z=(x_5,x_6)$  
 of the wavefunction  of the  KK modes\footnote{Let us show this for the KK zero modes. 
 The Wilson line changes the equation for the zero mode: 
$a\,\psi_0=0$:
$(i{\bar\partial}+ \frac{1}{2} i\, q_0\, g \,|f|z -q_0\, g\, \varphi^\dagger )\psi_0=0,$
with $z=x_5 +i x_6$.
Then, the solution for the zero mode becomes
$\psi_0=  h(z) \,\exp[-\frac{1}{2} \, q_0\, g \,|f|\, ({\bar z}-{\bar z}_0)(z-z_0)], $
where $z_0\equiv -2\sqrt 2 \,i \varphi^\dagger/|f|$ and $h(z)$ is a holomorphic function.  
Therefore, a constant Wilson line only  shifts $z$ by  $z_0$, but the number of 
zero modes (equivalent to the number of the 
possible centers within the fundamental domain on a torus) remains fixed by the 
quantisation condition as it is for a vanishing Wilson line. 
The same conclusion can be drawn in an asymmetric gauge for the background, such as
$A_5\!=\!0$ and $A_6\!=\!f x_5$.} 
 by an amount $\propto (\varphi_1/f, \varphi_2/f)$ and so the Wilson line does not  enter 
in the mass formulae of the  KK modes (and  of the potential).
This  explains  why the  momentum-independent correction $\delta m^2(0)$
 vanished at one-loop, with  $\varphi$  a flat  direction. This appears as
a consequence of the continuous (classical) translation symmetry of  $T_2$
which can  ``shift away'' $\varphi$, so  the KK spectrum  (and the potential) does not
depend on it.

The initial continuous translation symmetry of  $T_2$
is broken however at  the quantum level  by non-local Wilson loops\footnote{In orbifolds
(no flux) this translation symmetry is broken  explicitly
 by the orbifold fixed points.}. 
To see this, we put together the solution for the 
background gauge potential in the symmetric gauge and the constant Wilson line 
$\varphi\!=\varphi_1\!+\!i\varphi_2$ in the following 
form\footnote{The gauge  $A_5\!=-f x_6/2$,  $A_6\!=\!f x_5/2$ 
is not  invariant under translations $x_j\!\ra\! x_j+\delta_j$ ($j\!=\!5,6$), 
but a gauge transformation  $\vec A\!\ra \!\vec A\!-f/2 
\vec\nabla (\delta_5 x_6-\delta_6 x_5)$ with $\vec A\!=(A_5,A_6)$
restores the translation  symmetry.}:
$A_5=-\frac{1}{2}f x_6 + \varphi_2$ and 
$A_6=\frac{1}{2}f x_5 +\varphi_1$.
Now, the Wilson loops integrals
 are: $w_1(x_6)=\exp[i \,q_0 \,g \int_0^{a_5} d x_5 A_5]=\exp[i \,q_0\, g (- f x_6/2 +\varphi_2) a_5]$ and
  $w_2(x_5)=\exp[i\, q_0\, g \int_0^{a_6} d x_6 A_6]= \exp[i\,q_0\, g (f x_5/2 +\varphi_1) a_6]$ where 
$a_k=2\pi R_k$, $k=5,6$. 
The Wilson loops $w_{1,2}$ must be invariant, in particular under
 translations $x_6\ra x_6+ \delta_6$, $x_5\ra x_5+\delta_5$; this happens only if
 $\delta_5=4\pi k/(f a_6 q_0 g)= 2 a_5 k/N$ and $\delta_6=4\pi l/(f a_5 q_0 g)= 2 a_6 l/N$.
 Here $k, l, N\in \mathbb{Z}$ are integers and we used
the flux quantisation condition $f a_5 a_6 q_0 g=2\pi N$ (see Section 2).
As a result, the continuous translation symmetry of $T_2$  is broken 
 by  non-local  Wilson loops to a discrete (accidental) translation 
symmetry $x_5\!\ra\! x_5\!+\! a_5 (2 k/N)$, $x_6\!\ra\! x_6\!+\! a_6 (2l/N)$ \cite{t1,t2}.
With this continuous symmetry broken, one must investigate at higher orders if  
the one-loop flat direction of $\varphi$  can be maintained.

To complete our discussion, 
let us also examine what happens if the sum over the modes in the calculation of the quantum 
corrections to $\delta m_{b,f}(q^2)$ is truncated to
a fixed number of levels. Truncating the summation to $0\leq n\leq n_0-1$ for bosons ($n_0$ levels)
and to $0\leq n\leq n_0^\prime-1$ for fermions ($n_0^\prime$ levels) we find
from eqs.(\ref{b}), (\ref{f}), after some algebra\footnote{This
 is done by writing the ``truncated'' sum as a difference
of two infinite towers/sums, bringing in eq.(\ref{q1}) a difference of Hurwitz zeta functions for 
each zeta function present there, e.g. for bosons: $\zeta[\epsilon,\rho_1]\ra \zeta[\epsilon,\rho_1]
-\zeta[\epsilon, n_0+\rho_1]$ and similar for 
fermions with $\rho_1\ra \rho_2$, $n_0\ra n_0^\prime$. Similar 
for $\zeta[\epsilon-1,\rho_{1,2}]$.}
\bea
\delta m_b^2(q^2)&=&- \frac{1}{\epsilon} \,K_0\,\alpha\,n_0\, (2 n_0+1)+\cO(\epsilon^0)
\nonumber\\[-5pt]
\delta m_f^2(q^2)&=& \,\,\,\,
  \frac{1}{\epsilon}\,K_0\,\alpha\,n_0^\prime \,(2 n_0^\prime +q^2/\alpha)+\cO(\epsilon^0)
\eea
Their sum becomes, for $n_0=n_0^\prime$ (by supersymmetry)
\bea\label{tru}
\delta m^2(q^2)=-\frac{1}{\epsilon} \,K_0\,\alpha \,n_0\,(1-q^2/\alpha)+\cO(\epsilon^0)
\eea

\medskip\noindent
This shows that a  truncation of the towers to a same finite level would bring
 in the action a wavefunction renormalization for the superfield  $\varphi$  
 (due to the term  $\propto  q^2/\epsilon$) familiar in softly broken supersymmetry
and also a momentum-independent  quadratic divergence\footnote{This situation
 is worse than the  case of eq.16 in the first paper in  
\cite{G1} of ordinary orbifolds (no flux) 
where only usual $q^2/\epsilon$ poles i.e. wavefunction renormalization existed 
for a ``truncated'' tower summation. }
 $\propto \alpha/\epsilon$ (due to broken supersymmetry),
 but no higher dimensional counterterm is present. The theory is 
 exactly 4-dimensional and renormalizable.
Summing instead the whole tower, as we did, 
changes these two divergences into a {\it worse} ``quartic'' divergence $\propto q^4/(\alpha\epsilon)$
discussed earlier; this demanded instead a higher dimensional counterterm operator ($L$)
specific to non-renormalizable theories ($n_0, n_0^\prime$ being now infinite).

A situation similar to that above is expected for the  quantum corrections to the  gauge coupling
in this theory, when the 6D Lorentz invariance ``promotes''  counterterm 
(\ref{ct}) for the Wilson line to $F^{MN} \Box F_{MN}$ which also contains $F^{\mu\nu} \Box F_{\mu\nu}$.
This is  similar to 6D orbifolds without flux where such a higher dimensional counterterm 
($\cA \int\! d^2\theta\, \Tr W^\alpha\, \Box W_{\alpha}\!+\!\textrm{h.c.}$, in superfield notation)
 is generated \cite{G2} and is actually the reason for the so-called 
``power-like'' running near the compactification scale. 

\bigskip\smallskip
\section{Conclusions}

Compactifications of effective theories in the presence of magnetic flux are interesting
for model building since they provide  supersymmetry breaking and chiral fermions.
However, very few quantum calculations exist  in such cases and this motivated our study.
We examined the one-loop offshell correction to the two-point Green function of the Wilson line
self-energy, in 6D  N=1  Abelian gauge theories compactified on $T_2$ with  magnetic flux 
($\propto \alpha$). 
 The offshell
calculation is important and  {\it necessary} in order to  capture the usual effective theory
expansion in powers of $\partial/\Lambda$; (from a 4D view
$\Lambda\!\sim\! 1/\sqrt\cA\!\sim\!\sqrt\alpha$, $\cA$=torus  area).

Since the one-loop momentum integrals are UV divergent, a regularization
is needed. We used the DR scheme which preserves all the symmetries of the initial 6D gauge theory.
The result shows that in the limit of removing the regulator, the two-point Green function
has a pole which dictates the form of the counterterm.  
This is a higher dimensional (derivative) operator that was  also  seen  in 
 orbifold compactifications  without flux. 
One consequence of this
counterterm  is that a ghost state is present of  (mass)$^2\!\propto \alpha$. 
We showed that such  operator is equivalent to an operator of the  same dimension (six)
that is actually {\it polynomial} (quartic) in the charged superfields and is obtained 
by integrating out (decoupling) the ghost state.
This operator is enhanced by a larger compactification area (when effective theory is 
applicable) and a reminder that effective theories
are non-renormalizable.

After identifying the counterterm for the one-loop offshell self-energy,
one may also consider the  momentum independent
 mass correction $\delta m^2(q^2=0)$, which is the curvature of the corresponding one-loop 
potential. Unlike in orbifold compactifications (without flux), this mass correction
vanishes at one-loop and the Wilson line corresponds to a flat direction. 
The reason for this is  a  translation symmetry in internal dimensions which
is broken however at the quantum level by non-local Wilson loops.
It is worth investigating this issue beyond the one-loop order considered here.

\vspace{0.3cm}

\section*{Appendix}

In the text we used the following integrals in Euclidean space
\begin{eqnarray}
I_1\!\! & \equiv  &\!\! \int \frac{d^d p}{(2 \pi)^d}
\frac{p_\mu}{((p+q)^2+m_2^2)(p^2+m_1^2)}
 =  \frac{-q_\mu}{(4\pi)^{\frac{d}{2}}}
\int_0^1 d x \,x \, \Gamma\big[2-{d}/{2}\big]
\Big[ L(x,q^2,m_{1,2})\Big]^{\frac{d}{2}-2}
\nonumber\\
I_2\! \!& \equiv &\!\! \int 
\frac{d^d p}{(2 \pi)^d} \frac{1}{((p+q)^2+m_2^2)(p^2+m_1^2)}
 =\,  \frac{1}{(4\pi)^{\frac{d}{2}}}\,
\int_0^1 d x \, \,\, \Gamma\big[2-{d}/{2}\big]
\Big[ L(x,q^2,m_{1,2})\Big]^{\frac{d}{2}-2} 
 \nonumber\\
I_3\!\! & \equiv &\!\!  \int \frac{d^d p}{(2 \pi)^d} 
\frac{p_\mu p_\nu}{((p+q)^2+m_2^2)(p^2+m_1^2)}
 =  \frac{1}{(4\pi)^{\frac{d}{2}}}\,
\frac{\delta_{\mu\nu}}{2}
\int_0^1 \! d x \, \Gamma\big[1-{d}/{2}\big] 
\Big[ L(x,q^2,m_{1,2})\Big]^{\frac{d}{2}-1} 
\nonumber\\
&& \hspace{5.cm}
+\, \frac{1}{(4\pi)^{\frac{d}{2}}} \,\,q_\mu q_\nu\,\,
\! \! \int_0^1  d x \,  x^2\,
\Gamma\big[2-  {d}/{2}\big] 
\Big[ L(x,q^2,m_{1,2})\Big]^{\frac{d}{2}-2} 
\nonumber
\end{eqnarray}
where
\begin{equation}
L(x,q^2,m_{1,2})\equiv x \,(1-x)\, q^2 +x\,  m_2^2 +(1-x)
m_1^2
\end{equation}
and $\sum_\mu\delta_{\mu\mu}=d$, ($d=4-2\,\epsilon$).

\vspace{0.5cm}
\noindent
{\bf Acknowledgements:} 
The work of HML is supported in part by Basic Science Research Program through the National 
Research Foundation of Korea (NRF) funded by the Ministry of Education, 
Science and Technology (NRF-2016R1A2B4008759).

\end{document}